# Pressure Raman effects and internal stress in network glasses

*Fei Wang[1], S. Mamedov[2] and P. Boolchand*
Department of Electrical and Computer Engineering and Computer Science
University of Cincinnati, Cincinnati, OH 45221-0030

*B. Goodman*
Department of Physics, University of Cincinnati, Cincinnati, OH 45221-0011

*Meera Chandrasekhar*
Department of Physics, University of Missouri- Columbia, Missouri 65211

Raman scattering from binary $Ge_xSe_{1-x}$ glasses under hydrostatic pressure shows onset of a steady increase in the frequency of modes of corner-sharing $GeSe_4$ tetrahedral units when the external pressure P exceeds a threshold value $P_c$. The threshold pressure $P_c(x)$ decreases with x in the $0.15 < x < 0.20$ range, nearly vanishes in the $0.20 < x < 0.25$ range, and then increases in the $0.25 < x < 1/3$ range. These $P_c(x)$ trends closely track those in the non-reversing enthalpy, $\Delta H_{nr}(x)$, near glass transitions ($T_g$s), and in particular, both $\Delta H_{nr}(x)$ and $P_c(x)$ vanish in the reversibility window ($0.20 < x < 0.25$). It is suggested that $P_c$ provides a measure of stress at the Raman active units; and its vanishing in the reversibility window suggests that these units are part of an isostatically rigid backbone. Isostaticity also accounts for the non-aging behavior of glasses observed in the reversibility window.

## I. THREE ELASTIC PHASES IN NETWORK GLASSES

The nature of the glass transition continues to be a challenging issue in condensed matter science [1]. Multi-component network glasses exhibit striking trends with composition in certain region of connectedness [2] - as defined by their mean coordination number ($\bar{r}$). Chalcogens alloyed with Group IV and V elements of similar size and possessing a mean coordination number in the range $2 < \bar{r} < 3$ represent some of the best inorganic glass formers in nature. The general picture [3] is that, on quenching a glass forming liquid, structural arrest begins somewhat above the glass transition temperature $T_g$ and continued further freezing of local structures takes place below $T_g$. The resulting frozen-in variations in bond lengths of a given type that exist throughout the network of the glass, means that the structural backbone itself contains locally stressed regions. Relaxation [4] of these regions over time leads to aging effects and also to hysteretic behavior [2] upon thermal cycling across $T_g$.

Until recently, such relaxation was thought to be a universal property of a glass. However, in a region of optimal coordination ($\bar{r} \sim 2.4$), glasses have been found to behave differently from that expectation [2,5-9], in that their aging is greatly suppressed and the glass transition is thermally reversing in character. The value of $\bar{r} \sim 2.40$, the so-called mean-field value, is, to a first approximation, when the average number of local valence bond-stretching and bond-bending force constraints ($n_c$, Lagrangian constraints) on an atom coincides [10-12] with its translational degrees of freedom, $n_d = 3$. Recent experiments along with theoretical ideas suggest, more specifically, the existence of local and medium-range structures which are *isostatically* rigid, in that they satisfy $n_c = n_d$ on the atomic scale exactly, not just on the average. A more detailed picture [10-12] then emerges, namely, that there are three elastic phases in network glasses: a floppy or under-constrained phase when $\bar{r} < 2.40$ or $n_c < 3$; a stressed-rigid or over-constrained phase when $\bar{r} > 2.40$ or $n_c > 3$; and an intermediate or optimally constrained phase around $\bar{r} \sim 2.40$ (or $n_c = 3$) as schematically illustrated in Fig.1a. This means that there are *two*-elastic phase transitions in glasses as the count of Lagrangian constraints $n_c$ increases, the first one at $r_c(1)$, a transition between the floppy phase and the rigid, but stress-free, intermediate phase, and the second at $r_c(2)$, between the intermediate phase and the stressed-rigid phase. Only in strictly random networks, a rare circumstance, can the two transitions coalesce, giving rise to a solitary elastic phase transition as was predicted by J.C.Phillips[13] and M.F.Thorpe [14] in the early 1980s.(See Fig.1a.). The usual behavior observed in chalcogenide glasses is the opening of intermediate phases (IPs) between floppy and stressed-rigid elastic ones, the 'windows' as shown in Fig.1b. A number of properties of IPs have been studied in chalcogenide [5-12, 15-17, ] and chalcohalide glasses [18,19] . In the former, the width of the IP is of the order of $\Delta \bar{r} \sim 0.13$. This range of $\bar{r}$ is determined by the possibility of forming a stress-free backbone out of stoichiometrically different isostatically rigid local structures over a limited composition range [20]. For $Ge_xSe_{1-x}$ glasses, for example, such structures are chain segments of edge-sharing (ES) $Ge(Se_{1/2})_4$ and corner-sharing (CS) $GeSe_4$ tetrahedra. Numerical simulations [15] confirm these results and give a width of the IP that is in reasonable agreement with experimental results on binary $Ge_xSe_{1-x}$ and $Si_xSe_{1-x}$ glasses [9,16,17]. The IP in the two chalcohalide glasses, $Ge_{1/4}$ (Se or S)$_{3/4-y}I_y$ have been examined in detail [18,19] and are found to be rather narrow, $\Delta \bar{r} < 0.01$. The elastic phase boundary occurs near an iodine content y = 1/6, and it lies almost precisely at the mean field optimal constraint number $n_c = n_d$, we suppose because dangling ends, such as halogens, serve to cut characteristic rings where isostatic rigidity is nucleated. In consequence, there is only one chemical unit which is isostatic and nearly one composition where the backbone is composed predominantly of these units [18,19]. Even in this extreme case, the experimental evidence [19] points to the existence of a well defined but rather narrow IP.

The characterization of 'stress' in network glasses poses formidable issues both theoretically and experimentally. A particular challenge in this respect are the diamond anvil cell (DAV) pressure measurements of the behavior of Raman scattering mode frequencies in binary Ge-S and Ge-Se glasses made in the mid-1980s by K.Murase and Fukunaga [21,22]. They observed that the frequency of CS tetrahedral units near 200 cm$^{-1}$ blue-shifts as a function of hydrostatic pressure (P) only after P exceeds a threshold value $P_c$, which depends on the composition *x*. In the present work, we have confirmed the existence of the pressure thresholds in binary $Ge_xSe_{1-x}$ glasses and have established fairly comprehensive trends in $P_c(x)$. As x increases the threshold pressure $P_c(x)$ is found to decrease for 0.15 < x < 0.20, to nearly vanish in the IP phase, 0.20 < x < 0.25, and to increase again for 0.25 < x < 1/3. The vanishing of $P_c$ in the IP glasses, 0.20 < x < 0.25, is a *novel* feature for a glass because the behavior is characteristic of a crystalline solid.

The outline of the remainder of the paper is as follows: Section II describes the experimental details and presents the results on the present glass samples. In section III, we discuss the results, and comment on the distribution of stress in the three elastic phases of the present glasses. Section IV contains the summary and conclusions.

## II. EXPERIMENTAL PROCEDURE AND RESULTS

### A. Sample Synthesis.

Glasses were prepared [16,17] by reacting the 99.999% pure Ge and Se in evacuated ( 6 x $10^{-7}$ Torr) fused quartz ampoules by slowly heating up to at 950 ºC. Melts were homogenized for several days at 950ºC, and then the temperatures were lowered to 50 ºC above the liquidus to equilibrate melts for a few hours prior to a water quench. Freshly quenched glasses were stored in a dry ambient and allowed to age at room temperature for periods ranging from 2 to 52 weeks. Thermal and optical measurements were initiated after the freshly quenched samples had aged at least 2 weeks.

### B. Thermal characterization.

A model 2920 MDSC from TA Instruments Inc, operated at 3ºC/min scan rate and 1ºC/100s modulation rate was used [5-10] to study enthalpy input in the vicinity of the glass transition temperature $T_g$. Temperature modulated differential scanning calorimetry (MDSC) allows separating the total enthalpy flow into a thermally reversing component, that is, one which follows the modulated temperature variation, plus a non-reversing component. This separation is illustrated in Fig.2 for the case of a glass at x =

0.15. The former captures quasi-equilibrium thermodynamic properties of the metastable glass state, specifically its heat capacity jump, the inflexion point of which is used to determine $T_g$ [2,5-10]; while the latter component captures non-equilibrium effects including network configurational changes that occur upon softening of a glass. This component usually shows a peak as a precursor to $T_g$ with the integrated area under the peak ($\Delta H_{nr}$ : non-reversing enthalpy) serving as a quantitative measure of the hysteretic nature of the transition. For glasses in the intermediate phase near $\bar{r} \sim 2.4$, the $\Delta H_{nr}$ term nearly vanishes and the glass transitions acquire a thermally reversing character.

Figures 3a and b show the MDSC results for $T_g(x)$ and $\Delta H_{nr}(x)$, respectively, for a range of *x* that includes all three elastic phases. The latter quantity almost vanishes in the IP [16,17] but increases by an order of magnitude outside the IP, where it also shows significant changes under 'aging'; while $T_g$ changes relatively little in the same range of *x*. The MDSC results are included in Fig.3 for comparison with the Raman pressure data obtained in the present work which we present next.

### C. Raman Pressure Experiments.

Our DAC experiments used a Merrill-Bassett cell [23] with alcohol/methanol mixture (1:4) as a pressure transmitting medium and ruby chips as a monometer [24]. Raman scattering was excited using a 647.1 nm $Kr^+$ laser line and the scattered radiation was analyzed using a model T64000 triple monochromater system from Jobin Yvon Inc., a charged coupled device detector and a microscope attachment [7,16,17].

External stress compresses inter-atomic bonds and, in general, blue-shifts vibrational modes due to anharmonic effects. These pressure induced changes show striking differences between crystalline and disordered structures. Fig.4 shows selected Raman lineshapes observed at different values in P for (a) crystalline (α) - $GeSe_2$, (b) stoichiometric (stressed-rigid) $GeSe_2$ glass and (c) $Ge_{21}Se_{79}$ glass, which lies in the IP. In the crystal, the strongly excited 210 $cm^{-1}$ phonon ( related to the breathing mode of $Ge(Se_{1/2})_4$ tetrahedra) readily blue-shifts upon application of pressure, but in $GeSe_2$ glass, the corresponding vibrational mode near 200 $cm^{-1}$ does not shift until P exceeds 32 kbar [21]. An estimate of the inverse participation ratio [25] of the 200 $cm^{-1}$ mode suggests that it is localized over a few tetrahedral units in the glass. Fig. 5 summarizes the P- induced blue-shifts of the CS mode at different glass compositions x. These shifts are reversible in P and were deduced by deconvoluting the observed lineshapes [16,17] in terms of a superposition of Gaussians with unrestricted centroids, linewidths and intensities.

The central result of the present work is that the threshold pressure $P_c(x)$ vanishes in the IP, $0.20 < x < 0.25$, but increases monotonically as x moves away from this phase both at $x > 0.25$ and at $x < 0.20$. Trends in $P_c(x)$ and $\Delta H_{nr}(x)$ display striking similarities (Fig.3 b and c) in that both *vanish* in the IP. Compositional trends in the fractional blue-shift $(dln\nu/dP)$ of the CS mode with mode frequency ($\nu_{CS}$) at $P > P_c$ are summarized in Fig. 6(b). We find that the fractional response steadily decreases as x increases. These results can be put in perspective by plotting them on a universal plot of the response as a function of mode frequency as illustrated in Fig. 6(a). In Fig.6(a) the data points of c-$As_2S_3$ are taken from the work of B.Weinstein and R.Zallen [26,27], the data points on α-$GeSe_2$ and Ge-Se glasses are taken from the present work. In general, as the mode frequency increases the fractional response decreases because the restoring forces associated with higher mode frequencies are larger, being more covalent in nature.

### III. DISCUSSION

#### A. $Ge_xSe_{1-x}$ Glass Molecular Structure, Onset of Rigidity and Pressure effects

The picture of glass structure evolving when Ge is alloyed in a Se base glass is that chains of $Se_n$ are *stochastically* cross-linked by Ge atoms [28] to form CS $GeSe_4$ tetrahedral units at low x, i.e., in the $0 < x < 0.10$ range. These tetrahedra are isolated in $Se_n$ chains, but with increasing x the relative spacing between these tetrahedra decreases and some edge-sharing (ES) tetrahedra begin to emerge [16,17] as x increases to 0.10. Variations of $T_g(x)$ in the low x range yield a slope ($dT_g/dx$ = 4.5°C/at.%Ge) that is in excellent accord with the parameter free prediction ($dT_g/dx$ = $T_o/\ln 4/2$) of this slope based on stochastic agglomeration theory [29,30] SAT. Here $T_o$ = 313K or 40° C and represents the $T_g$ of the base glass of pure Se. SAT has proved to be a powerful method to understand the connection between glass molecular structure and $T_g$. The structural interpretation of $T_g$ suggested is that it measures the connectedness of the network backbone. At higher x ($> 0.10$), three distinct types of local structures are formed and include, $Se_n$ chains, ES- and CS-tetrahedra. One then expects a superlinear variation of $T_g(x)$ to be manifested as more local structures appear in the backbone. A second order elastic phase transition, viz., the rigidity transition, occurs when x increases to 0.20 or $r = r_c(1) = 2.40$ as isostatically rigid local structures percolate. In the $0.20 < x < 0.25$ range, the backbone consists predominantly of two isostatically rigid local structures : CS $GeSe_4$ units and ES $GeSe_2$ units. The IP ambient pressure Raman scattering reveals that the optical elasticity ( $v_{CS}^2(x)$) displays a power-law p = 0.75(15). A first order elastic phase transition, viz. stress transition, occurs when x increases to 0.26 or $r = r_c(2) = 2.52$. The optical elasticity shows a first order jump at the phase boundary followed by a power-law behavior in the stress-rigid phase ($0.26 < x < 1/3$) of p = 1.54(10). The latter result is in excellent accord with numerical simulations [31,32] based on the standard model of glasses as random networks. With increasing x, a global maximum in $T_g$ sets in near x = 1/3. Here again SAT provides the most natural interpretation of the result in suggesting that the global connectivity of the backbone is compromised as nano-scale phase separation (nsps) [33] sets in. In particular, both Raman and Mossbauer spectroscopy provide evidence for growth of Ge-rich ethanelike clusters ($Ge_2Se_6$) once x > 0.31. These new clusters most likely nsps [34] from the backbone, because in the thermal measurement one observes a drastic reduction in the slope, $dT_g/dx$ , exactly at the same composition, i.e. at x > 0.31. Near x ~ 1/3, the slope $dT_g/dx$ vanishes understandably because $T_g$ has a global maximum.

Our Raman pressure measurements show a systematic increase in the scattering strength ratio of the CS to ES mode ( $A_{CS}/A_{ES}(x,P)$ ) as a function of P at glass composition x in the $0.25 < x < 1/3$ range. Fig. 7 gives an overview of the variations in the $A_{CS}/A_{ES}(x,P)$ ratio. It would be relevant to recall here that as $x > 0.25$, the CS Ge(Se$_{1/2}$)$_4$ units become overconstrained ($n_c = 3.67$) while ES Ge(Se$_{1/2}$)$_4$ units remain optimally constrained ($n_c = 3.0$). We interpret this increase as showing a shift in concentration from ES units to CS units as illustrated, for example, in Fig 4 for GeSe$_2$ glass. If the energies of local structures containing an ES unit or a CS unit are raised under pressure from their $P = 0$ values by an amount $\Delta\varepsilon_{cs}(P)$ and $\Delta\varepsilon_{es}(P)$, respectively, the fractional concentration of these units will be given by the Boltzmann factor,

$$f = \exp[(\Delta\varepsilon_{es}(P) - \Delta\varepsilon_{cs}(P))/T] \qquad (1)$$

with $\Delta\varepsilon_{es}(P) > \Delta\varepsilon_{cs}(P)$. To obtain an estimate of the energy change we assume that the local structures have volume $V_E$ and $V_C$, respectively, and that each is of order $V_T$, the volume of a Ge(Se$_{1/2}$)$_4$ tetrahedron, say, $V_C = aV_T$ and $V_E = bV_T$. Because $V_E$ contains two tetrahedral units (Fig. 9) we expect that $b > a$. Applying the macroscopic elastic energy formula,

$$\Delta\varepsilon_s(P) = P^2 V / 2K \qquad (2)$$

with K representing the bulk modulus, one obtains the fraction of ES/CS units,

$$f \sim \exp[(b - a)\tau(P)/T] \qquad (3)$$

where $\tau(P) = P^2 V_T/2K k_B$. Taking the value of the bulk modulus from the measured sound velocities [35] at these compositions, gives $K \approx 280$ kbar. Taking $V_T = 1.03$ $d^3$ where d is the Ge-Se distance, i.e., the sum of the ionic radii, $\cong 2.4$ A, gives $\tau(P) \approx 1.7(P/kbar)^2 k_B$. For example, at room temperature and P = 10kbar, $f \sim \exp\{0.57(b - a)\}$; and comparison with Fig. 8a gives the estimate: $V_E - V_C \approx 0.44$ $d^3$. This simplistic numerical exercise provides some *plausibility* to the strain-energy picture in that it gives the right pressure trend. It must not be pushed too far however. The proportionality $\Delta\varepsilon(P) \propto P^2$ for a microscopic linear continuum is not exactly supported by the results of Figs. 7a and 8a. The challenges in understanding issues like these in glasses are formidable. One need only be reminded of the two-level theory of low temperature specific heat of amorphous solids [36] which still lacks an agreed structural picture.

Raman linewidths display features related to the rigidity of the crosslinking tetrahedra in the backbone. The linewidth ($\Gamma = 15$ cm$^{-1}$) of the CS mode at x = 0.25 increases steadily with x to reach $\Gamma = 16$ cm$^{-1}$ at x = 0.30. GeSe$_4$ units, which are optimally constrained at $n_c = 3$ (Ref.20), predominantly populate the IP backbone; while overconstrained Ge(Se$_{1/2}$)$_4$ units ($n_c = 3.67$) are largely present in stressed-rigid glasses near x = 1/3. Redundant bonds in the latter units result in larger distortions of Ge-Se bond-lengths and Se-Ge-Se bond-angles and contribute to the broadening of the CS-mode with x. The second observation is that ES modes have a narrower $\Gamma$ than CS modes in the IP (12 cm$^{-1}$ vs 15 cm$^{-1}$) even though both units are optimally constrained (n = 3). In a CS unit all of the Ge-Se bonds connect to the backbone as against only 2 bonds in ES units (Fig. 9); so that the latter tetrahedra are more decoupled from stress-induced deformations of the backbone which contribute to the linewidth.

Murase and Fukunaga [21] were the first to suggest that $P_c$ provides a measure of an internal pressure (stress) which must be exceeded by the applied pressure before vibrational modes blue-shift. Redundant bonds in the stressed-rigid phase put the rest of the backbone under a large stress, and as expected, $P_c$ steadily increases (Fig 1c) with the concentration of such bonds as x increases to 1/3. The mechanical equilibrium prevailing in the IP is also disrupted in floppy-glasses [16] as evidenced by $P_c$ increasing as x < 0.20. In the floppy phase the fraction of polymeric $Se_n$ chain fragments increases as x→0. These fragments are intrinsically underconstrained ($n_c$ = 2) [20], and exert an entropic pressure on the cross-linked segments.

**B. Stress in Network Glasses- Theoretical considerations**

The interpretation [21] that $P_c$ provides a measure of the internal stress which must be exceeded by the applied pressure is a phenomenological one. Obviously, especially for a disordered network, the atomic-scale distribution of bond bending and bond stretching cannot be represented [37] by $P_c$, nor by a 6-component stress tensor – as it could be for a Bravais crystalline lattice. More importantly, the suggestion that $P_c$ is somehow related to a residual internal stress like, say, from quenching, cannot explain threshold behavior, which is nonlinear. In a linear elastic system stresses from external forces add linearly to internal stresses – as long as elastic limits are not exceeded.

Furthermore, while the CS mode central frequency does not shift for $P < P_c$, the mode linewidth (Fig. 8b) increases over 40% in that pressure range.

Even in the IP where $P_c$ vanishes the linewidth is seen (Fig. 4) to be more sensitive to external pressure than is the change in the CS mode central frequency. This probably has its origin in the relatively weaker bond-bending forces in relation to bond-stretching forces that allow larger bond angle distortions relative to bond length changes. The importance of bond-bending interactions [38] is central in understanding the glass forming tendency of covalent systems is central. Angular distortions mix the CS modes of a particular $Ge(Se_{1/2})_4$ tetrahedron with modes of different symmetry (including Raman inactive modes) of higher and lower frequencies than $\nu_{CS}$, thereby shifting $\nu_{CS}$ either up or down depending, on the particular local distortion, and producing an inhomogeneous line broadening.

The question remains: What kinds of distortions in an inhomogeneous network are capable of producing a nonlinear elasticity threshold? One possibility is internal contact-like behavior in which both rigid structures and weaker structures coexist but with weak effective contact between them. Initially, the rigid subsystem provides a buffer against the external pressure; but eventually a strong contact develops in a highly nonlinear expansion of the contact area, and further pressure increase is transmitted more or less homogenously throughout the entire system.

**C. Stress-free Nature of the Intermediate Phase**.

The IP (0.20 < x < 0.25), though comprising disordered networks, has some similarity to crystals. Both form space filling networks that are homogeneous at a mesoscopic level. The space filling property of the IP backbones is revealed by a minimum in their molar volumes [39]. In glasses, the local and medium range structures prevailing in the IP are isostatically rigid, with no redundant bonds [10,15,18] to produce stress. Therefore, IP's are *not* made up of segregated regions of different elasticity. This picture of relative homogeneity of stress distribution at intermediate ranges is consistent with the fact (Fig.3) that $P_c$ is zero in the otherwise quite different systems, α-GeSe$_2$ and the glass IP. These observations suggest that absence of a pressure threshold as well as absence of appreciable aging effects are part of a broader set of consequences of IP glasses being in a state of metastable *mechanical-equilibrium*; which itself derives from the optimal network connectedness of having *atomically exact* constraint balance $n_c = n_d$. In contrast, we might better think of $P_c$ as being a measure of intermediate range stress *inhomogeneity* rather than as a macroscopic residual stress.

### D. Atomic size and stress distribution in network glasses

Our choice of the Ge-Se binary in the present study deserves a final comment. The Ge-Se binary is an ideal system for investigations of network stress in a disordered system by Raman scattering. First, the three elastic phases in this binary are well documented [16]. Second, local stress compensation effects due to atomic size mismatch are minimized because the atomic size of Ge (1.22 A) and Se (1.17A) are nearly the same. This has the important consequence that stress is globally distributed, and the frequency of the Raman active CS mode can serve as a good representation of backbone stress. When constituent atoms have different sizes however, local stress relief can occur, and preclude a reliable measurement of the backbone stress from a measurement of a Raman mode frequency alone.

The above ideas are strongly supported by the DAC measurements on corresponding binary $Ge_xS_{1-x}$ glasses. Fig. 10 shows the pressure variations in mode frequency of the CS $Ge(S_{1/2})_4$ tetrahedral units at x = 15%, 23% and 33% in panels (a) , (b) and (c) respectively. These compositions belong respectively to the floppy (15%), intermediate (23%) and stressed-rigid phases (33%) of binary Ge-S glasses. The behavior seen in Fig. 7 is quite different from the systematic patterns of Fig. 5, which we take to support the previously mentioned role of relative atomic sizes. The S anion size of 1.02 A is significantly smaller than that of the Ge cation. A pattern of stress relief by the tetrahedral units distorting locally should be reflected in the distribution of mode frequencies from place to place in the network that contributes to an inhomogeneous broadening of the CS mode in the Raman scattering. Indeed, we find that the linewidth (FWHM) of the CS mode in binary $Ge_xS_{1-x}$ glasses are nearly *twice as large* as in corresponding selenide glasses. These results suggest that Raman active CS mode frequency as a probe of network stress in glassy networks can only be relied upon when the atomic size of the network forming atoms are nearly the same. Thus, glass systems based on the Ge-As-Se, or the Si-P-S ternaries, or the P-S binary would appear to be ideally suited for Raman Pressure measurements.

## IV. CONCLUSIONS

In conclusion, pressure effects in Raman scattering on the $Ge_xSe_{1-x}$ binary highlight a new feature of the IP ($0.20 < x < 0.25$) ; threshold-pressures ($P_c$) above which modes blue-shift are found to *vanish* in this phase. The behavior ( $P_c = 0$) is characteristic of a crystalline solid and probably of window glass as well [40], and suggests that glassy networks in IPs are present in a state of mechanical equilibrium that is connectivity driven. Glass compositions in the IP are viewed to be self-organized, and the non-aging of the non-reversing enthalpy ($\Delta H_{nr}$) in this phase appears to be a consequence of that equilibrium. Non-aging of self-organized disordered networks is a novel basic idea that has profound technology implications including the design of a new generation of thin-film gate dielectrics [41].

**Acknowledgements**. We thank S. V. Sinogeikin for assistance with the high-P experiments. This work is supported by NSF grants DMR- 01-01808 and 04-56472.

[1] Present address: Department of Electrical Engineering, California Polytechnic State University, San Luis Obispo, CA 93407.

[2.] Raman and X-ray Fluorescence Group, HORIBA Jobin Yvon Inc., 3880 Park Ave, Edison, NJ 08820-3012.

Figure Captions

Fig.1.(a) Schematic of the three elastic phases observed in network glasses as a function of increasing connectivity. In select cases of truly random networks the intermediate phase collapses yielding a solitary elastic phase transition from a floppy to a stressed-rigid. (b) Observed intermediate phases or reversibility windows in indicated binary and ternary glasses. In the two chalcohalide glasses $Ge_{1/4}(S \text{ or } Se)_{3/4-y}I_y$ , note that the intermediate phase almost totally collapses.

Fig.2. T-modulated DSC scan of a $Ge_{15}Se_{85}$ glass showing a $T_g$ = 129.7(1.0)°C, inferred from the inflexion point of the reversing heat flow signal. Note a sizable non-reversing heat flow , $\Delta H_{nr}$ ( shaded area), for this floppy glass composition. The scan rate was 3°C/min and the modulation rate was 1°C/100s.

Fig. 3. Summary of MDSC results on Ge-Se glasses showing variations in (a) glass transition temperature $T_g(x)$ and (b) non-reversing enthalpy near $T_g$ $\Delta H_{nr}(x)$. DAC results on variations in Raman pressure thresholds $P_c(x)$ is plotted in (c). The (□) gives the $P_c$ result on α-$GeSe_2$. Note $\Delta H_{nr}(x)$ and $P_c(x)$ both vanish in the IP.

Fig.4. Raman lineshapes observed as a function of P in (a) α-$GeSe_2$ (b) $GeSe_2$ glass and (c) $Ge_{21}Se_{79}$ glass. Note $P_c$ = 0 in $Ge_{21}Se_{79}$ glass, a composition in the intermediate phase, but not in $GeSe_2$ glass ( stressed-rigid).

Fig. 5. Variations in the frequency of CS tetrahedral units as a function of P for select $Ge_xSe_{1-x}$ glasses studied in the present work (●). The results show the existence of pressure thresholds ($P_c$) in the stressed-rigid (25%, 30% and 33.33%) and floppy (15%, 17%) glass compositions but not in the IP (20%, 22%) ones. The At x = 15%, a two- line fit (as shown) not only yields a lower chi-square than a one-line fit but also yields a slope (dν/dP) that is nearly 65% larger and forms part of a general trend as as x- or ν- decreases as sketched in Fig. 6. The ( Δ ) data points are taken from the work of Murase and Fukunaga [21.

Fig. 6. (a) Fractional blue shift of the Raman vibrational modes in c-$As_2S_3$(●) [27], α-$GeSe_2$ (□)and present $Ge_xSe_{1-x}$ glasses ( ▲ ) plotted as a function of mode frequency. The steady decline in the response with increasing frequency reflects the higher restoring force associated with the covalent interations in relation to the van der Waals ones. (b) Fractional blue shift of the CS mode frequency showing a steady decrease with glass Ge concentration x reflecting the increased rigidity of the network with increasing connectivity.

Fig. 7. Pressure-induced changes in the (a) CS/ES fraction (b) FWHM of CS mode and (c) FWHM of ES mode for a $Ge_{25}Se_{75}$ glass composition.

Fig. 8. Pressure-induced changes in the (a) CS/ES fraction (b) FWHM of CS mode and (c) FWHM of ES mode for a $GeSe_2$ glass composition.

Fig. 9. Schematic drawing of (a) CS and (b) ES units. The 4-fold coordinated atom is Ge while the 2-fold coordinated one is Se.

Fig. 10. Pressure induced changes in CS mode frequency of $Ge_xS_{1-x}$ glasses at (a) x = 15%, (b) 23% and (c) 33% showing that $P_c \sim 0$ in all three glass compositions. See text for details.

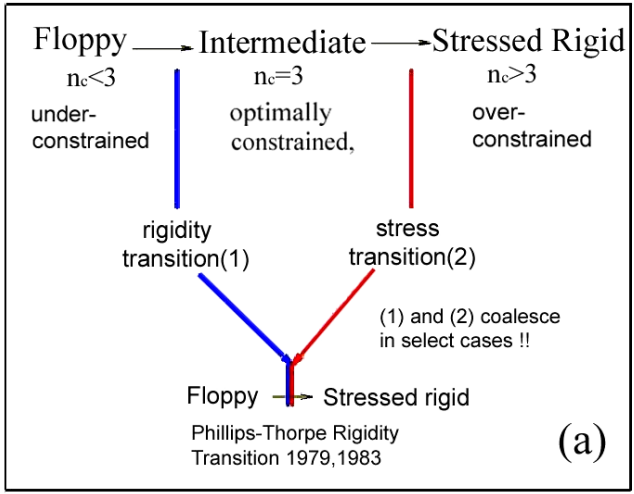

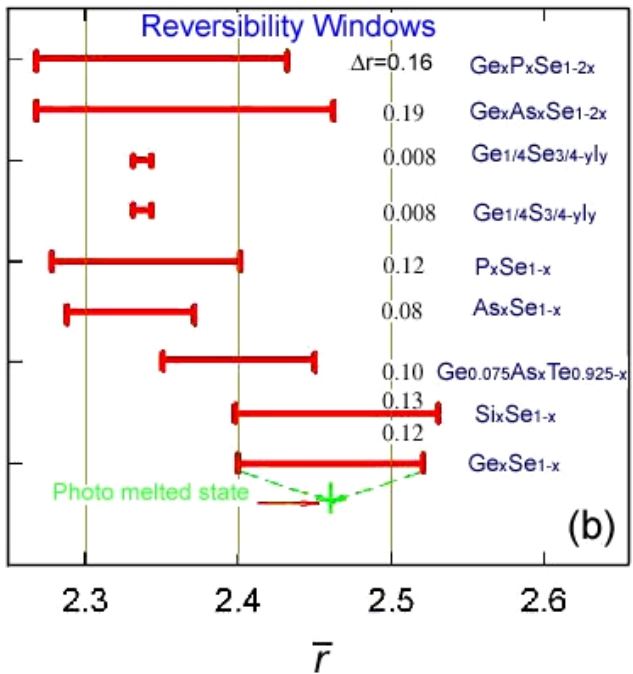

Figure 1.

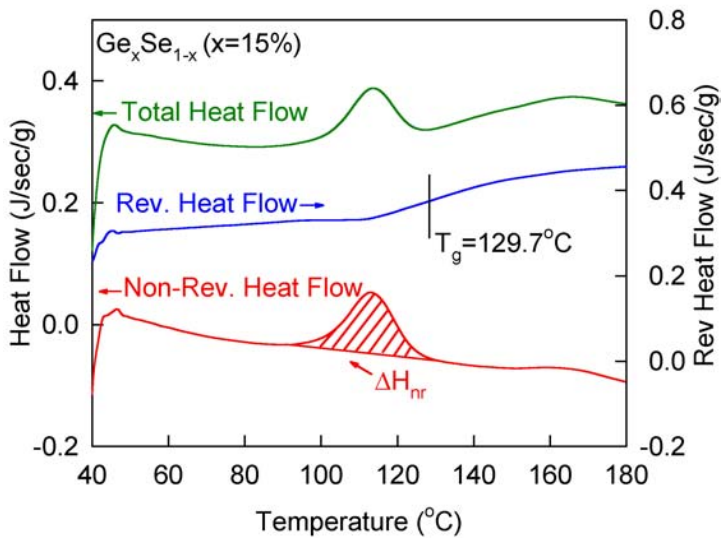

Figure 2.

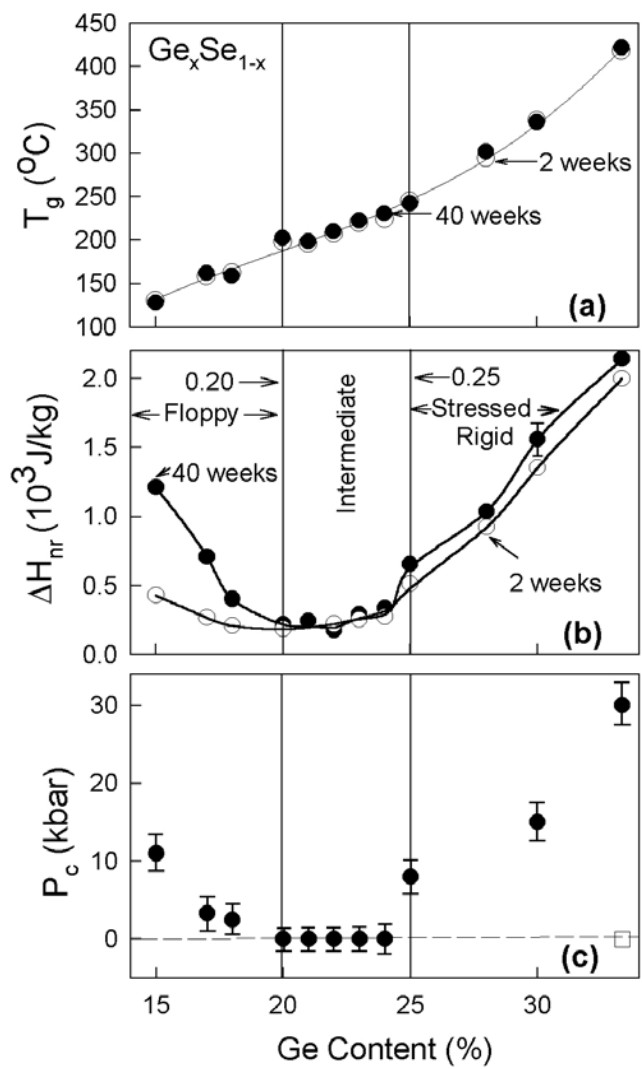

Figure 3.

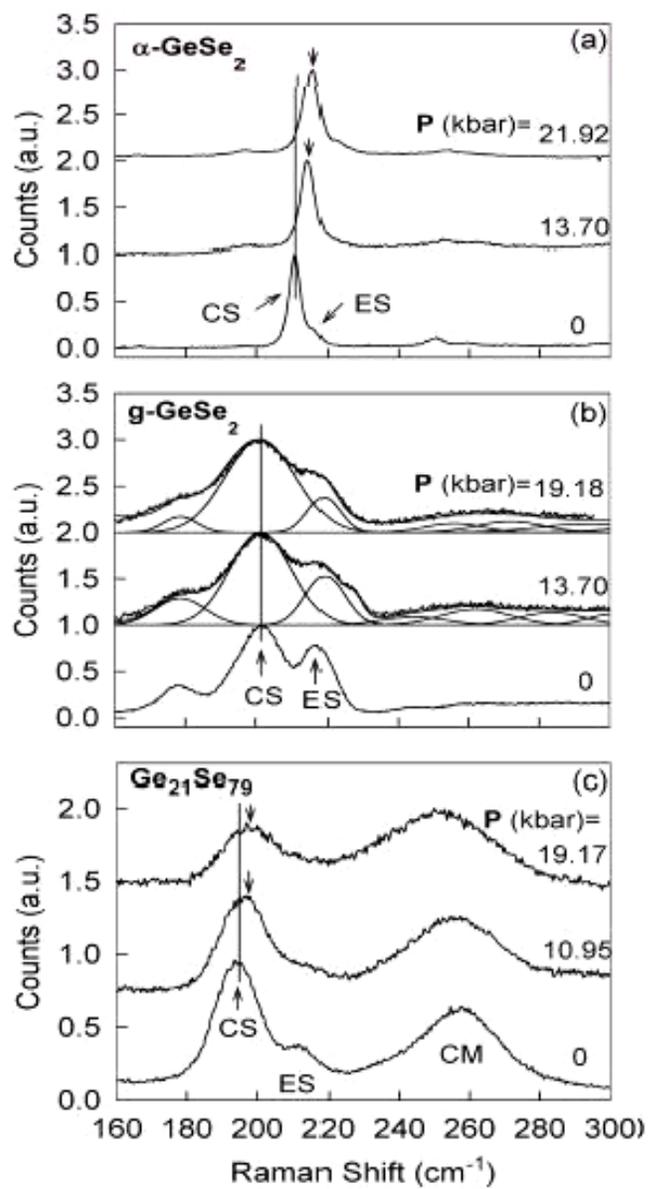

Figure 4.

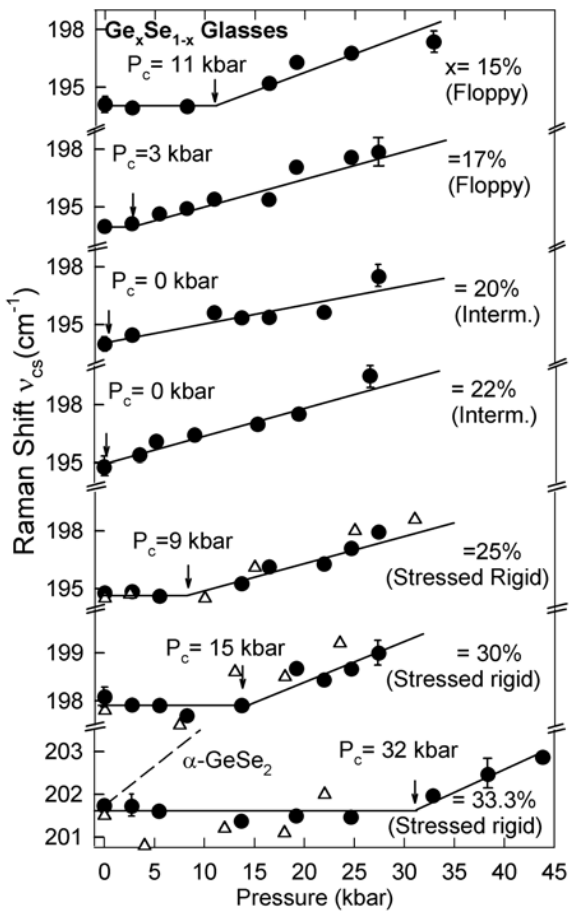

Figure 5.

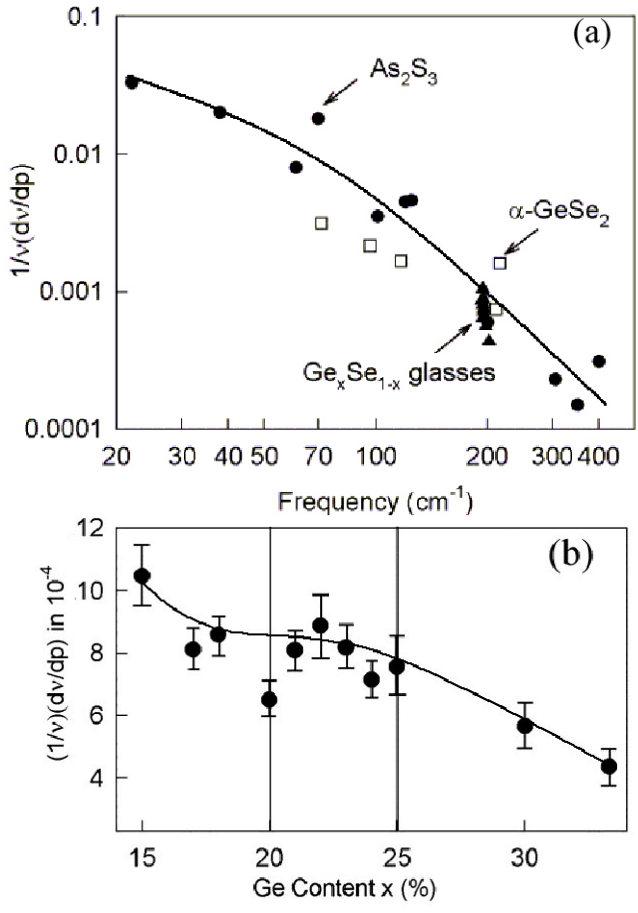

Figure 6.

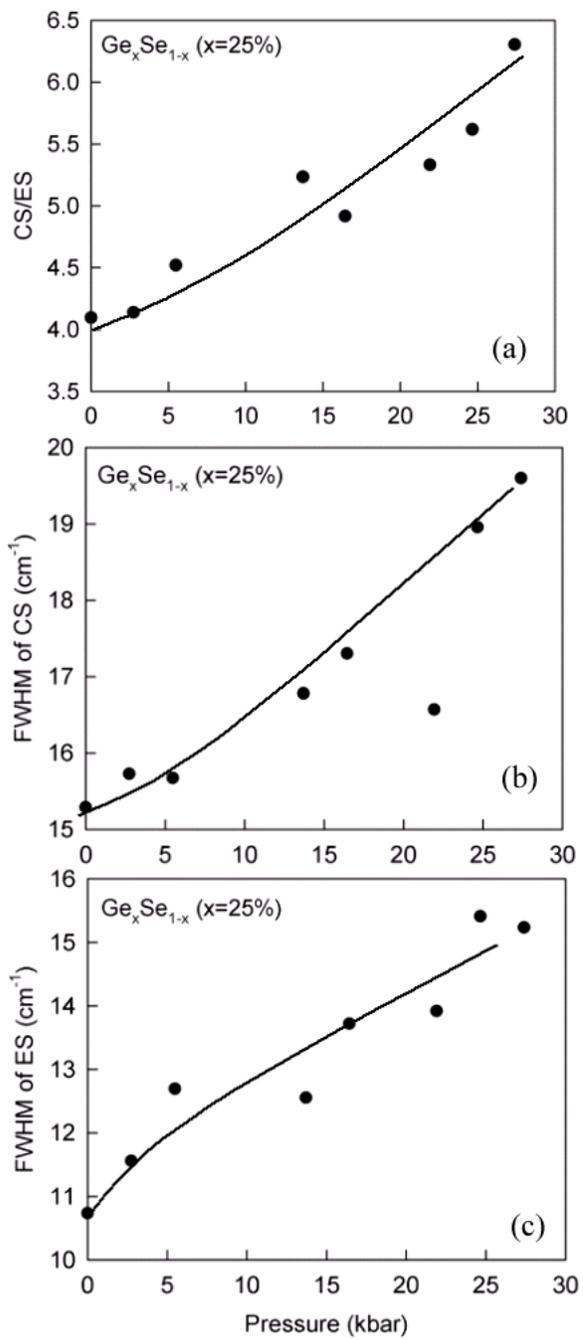

Figure 7.

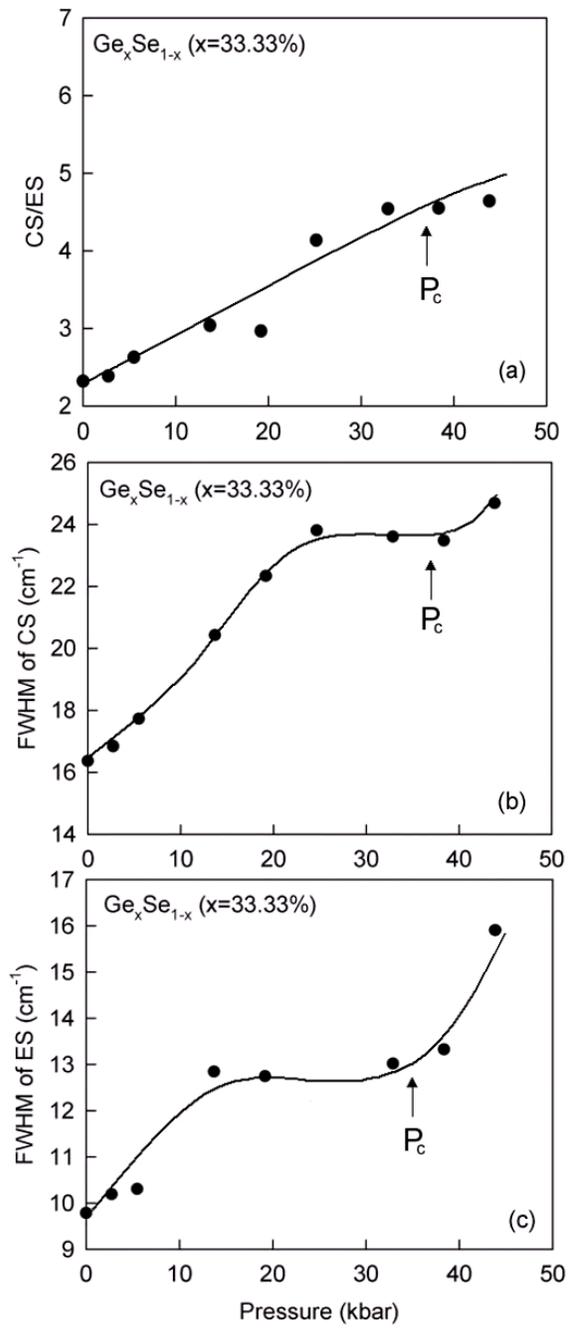

Figure 8.

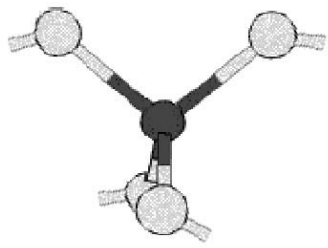

CS Ge(Se$_{1/2}$)$_4$ unit

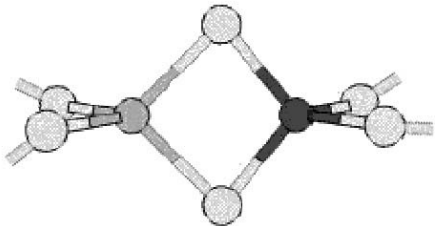

ES GeSe$_2$ unit

Figure 9.

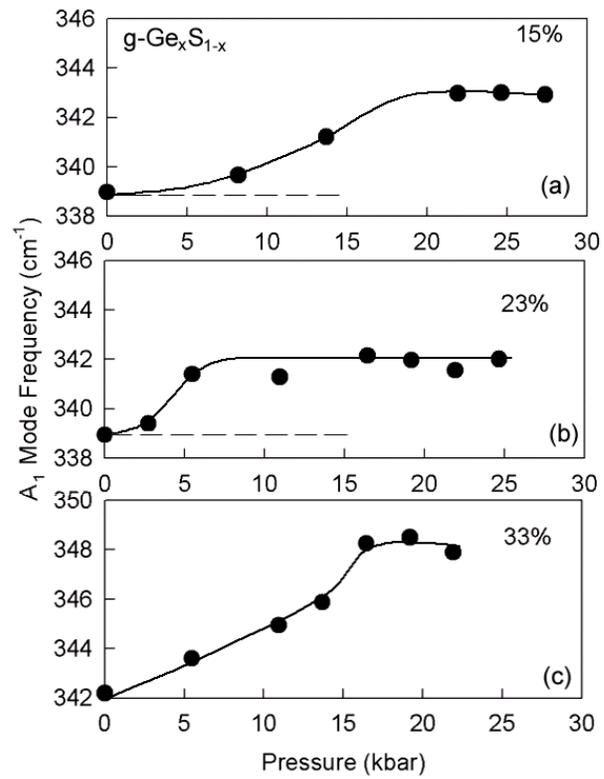

Figure 10.